\documentclass[floatfix, noshowpacs, preprintnumbers, twocolumn, nofootinbib, aps, pra,10pt]{revtex4-1}

\usepackage{amsmath,amsthm,amssymb,amsfonts,setspace}
\usepackage{bm}
\usepackage{bbold}
\usepackage{float}
\usepackage{graphicx}
\usepackage{mathtools}
\usepackage[colorlinks]{hyperref}
\usepackage[caption=false]{subfig}
\usepackage{url}

\mathtoolsset{centercolon}

\newcommand{\bra}[1]{\left\langle{#1}\right\vert}
\newcommand{\ket}[1]{\left\vert{#1}\right\rangle}
\newcommand{\eq}[1]{Eq.~\hyperref[eq:#1]{(\ref*{eq:#1})}}
\newcommand{\eqbeg}[1]{Equation~\hyperref[eq:#1]{(\ref*{eq:#1})}}
\newcommand{\eqs}[2]{Eqs.~\hyperref[eq:#1]{(\ref*{eq:#1})} and \hyperref[eq:#2]{(\ref*{eq:#2})}}
\renewcommand{\sec}[1]{\hyperref[sec:#1]{Section~\ref*{sec:#1}}}
\newcommand{\fig}[1]{\hyperref[fig:#1]{Fig.~\ref*{fig:#1}}}
\newcommand{\sfig}[2]{\hyperref[fig:#1]{Fig.~\ref*{fig:#1}#2}}
\newcommand{\thm}[1]{\hyperref[thm:#1]{Theorem~\ref*{thm:#1}}}

\newcommand{\cnot}{\textsc{cnot}}
\newcommand{\cz}{\textsc{cz}}

\newtheorem{theorem}{Theorem}

\theoremstyle{definition}

\newtheorem*{definition*}{Definition}

\begin{document}
\title{The complexity of simulating constant-depth BosonSampling}

\author{Daniel J.\ Brod}
\email{dbrod@perimeterinstitute.ca}
\affiliation{Perimeter Institute for Theoretical Physics}
\date{\today}

\begin{abstract}
BosonSampling is a restricted model of quantum computation proposed recently, where a non-adaptive linear-optical network is used to solve a sampling problem that seems to be hard for classical computers. Here we show that, even if the linear-optical network has a constant number (greater than four) of beam splitter layers, the exact version of the BosonSampling problem is still classically hard, unless the polynomial hierarchy collapses to its third level. This is based on similar result known for constant-depth quantum circuits and circuits of 2-local commuting gates (IQP).
\end{abstract}

\maketitle

\section{Introduction} \label{sec:intro}

Recently, several restricted models of quantum computation have been investigated that seem to lie between classical and quantum, in terms of their computational complexity. While unlikely to be universal for quantum computation, these models seem to have some nontrivial computational power, as they were shown to perform tasks considered hard for classical computers. Unlike factoring, the tasks performed by these restricted models do not seem to be in NP nor to be related to ``useful'' applications, making them less appealing from a practical point of view than the more standard applications of quantum computers, such as Shor's factoring algorithm \cite{Shor1997}. On the other hand, it is conceivable that these models are easier to implement in experimental settings, since they do not require the full power of quantum computation, and thus provide good candidates for intermediate milestones for the field. Examples of these models include the one clean qubit model and generalizations \cite{Knill1998, Morimae2014a,Morimae2014b}, quantum circuits of constant depth three or greater \cite{Terhal2004}, circuits of commuting 2-local quantum gates, also known as IQP \cite{Bremner2011}, non-adaptive measurement-based quantum computation, known as MBCC (measurement-based classical computation) \cite{Hoban2014}, and non-adaptive linear optics, also known as BosonSampling \cite{Aaronson2013a}.

However, even if these restricted models propose demonstrations closer to the reach of current technologies than universal quantum computers, there are still important technological challenges to overcome before we can provide compelling evidence that quantum systems can outperform classical computers. BosonSampling, for example, is in a sense more robust than the other models mentioned since its complexity claims hold even in an approximate setting, thus accommodating (some level of) experimental imperfections, and an experiment demonstrating interference of a few dozen photons in an interferometer of a few hundred modes already would provide evidence of the computational power of quantum systems long before any reasonable-sized implementation of Shor's algorithm. Even in this arguably more ``experimental-friendly'' scenario, the best experiments reported so far all lie in the limited range of 3-4 photons \cite{Broome2013, Crespi2013b,Spring2013,Tillmann2013,Spagnolo2013b,Spagnolo2014,Carolan2014}. This motivates the study of further simplifications of these models, both from the practical point of view of bringing them even closer to experimental reality, as well as from a more conceptual point of view of understanding the origins of their complexity.

In this paper, our main focus is on a restricted version of the BosonSampling model where the linear-optical circuits have only a constant number of layers of beam splitters, which we call constant-depth BosonSampling. Using techniques similar to those used for constant-depth quantum circuits \cite{Terhal2004} and IQP \cite{Bremner2011} we will prove that, if it was possible to efficiently produce a sample from the \emph{exact} output distribution of a linear optical experiment of depth four (i.e.\ with four sequential layers of beam splitters) by only classical means, this would imply a large collapse in the structure of complexity classes (namely, the polynomial hierarchy would collapse to its third level), which is considered a highly unlikely result. Along the way we will construct a simple example of a circuit that is both in IQP and has depth four, thus at the intersection of two of the aforementioned restricted models---while this has been noticed before, we believe the presentation provided here may be of independent pedagogical interest.

Optimizing the BosonSampling model in terms of depth is of special interest for recent experimental implementations, as several use an integrated approach to linear optics where interferometers consist of waveguides etched into e.g.\ glass chips \cite{Broome2013, Crespi2013b,Spring2013,Tillmann2013,Spagnolo2013b,Spagnolo2014,Carolan2014}. While having the benefit of miniaturization, allowing for implementation of networks of several dozen beam splitters within a few centimeters of material, these approaches also often suffer from certain scalability issues---in particular, one can model losses by attributing a certain loss probability to each beam splitter a photon must transverse inside the chip, which leads to a decay of the signal that is exponential in the circuit depth. While further technological advances may mitigate this issue, this clearly motivates an investigation of how small the depth of an optical circuit needs to be to allow for some nontrivial complexity claim. In Ref.\ \cite{Aaronson2013a}, the authors show than an $n$-photon $m$-mode BosonSampling device can be parallelized to a circuit of O($n$ log $m$) depth even in the approximate setting. Our result shows that a depth of four is in fact already sufficient in the \emph{exact} setting, and this result may be a first step in reducing the depth requirements of the original model.

This paper is organized as follows. In \sec{background} we give some basic definitions and theoretical background. In \sec{postselection} we discuss some relations between postselection and complexity classes relevant for this work, outlining the main complexity argument leading to our result. In \sec{KLM} we give a brief overview of the KLM scheme for quantum computing with linear optics. In \sec{cdepthQC} we give a simplified proof of the (previously known) fact that an efficient classical algorithm for sampling from the exact output distribution of a depth-d quantum circuit would imply the collapse of the polynomial hierarchy to its third level, for any $d>3$. In \sec{cdepthLOQC} we show how the result of \sec{cdepthQC} can be adapted to the setting of linear-optical circuits of depth four. Finally, \sec{conclusion} contains concluding remarks and remaining open questions.

\section{Preliminary definitions} \label{sec:background}

\subsection{Postselection and complexity} \label{sec:postselection}

Let us begin with a brief outline of the main complexity-theoretical argument behind our result. This reasoning is a simplified version of that used, e.g., in Refs.\ \cite{Bremner2011,Aaronson2013a} to prove similar results for different models.

Let A be some model of probabilistic computation. It may correspond to probabilistic classical computation (BPP), quantum computation (BQP), or some other restricted model such as circuits of commuting quantum gates (IQP) of Ref.\ \cite{Bremner2011}, quantum circuits of constant depth \cite{Terhal2004}, or linear optical circuits (LO) such as those used for BosonSampling \cite{Aaronson2013a}. For all of these cases, we can define a corresponding complexity class postA, or ``A with postselection''. Informally, consider a family of circuits from the restricted model A which have two registers, the output register ($o$) which encodes the answer to some desired problem, and the postselection register ($p$) which flags whether the computation succeeded or not. The class postA then corresponds to the set of problems for which there exists a family of circuits in A such that (i) the probability that the computation succeeds (i.e.\ the probability of observing a specific predetermined outcome on register $p$) is nonzero and (ii) conditioned on this success, the circuit solves the desired problem with probability $1-\epsilon$ for some arbitrary positive constant $\epsilon < 1/2$. For our purposes, this informal definition will suffice, but a more formal definition of these classes can be found in Refs.\ \cite{Aaronson2013a,Bremner2011}.

Two subtleties in the definition of postA should be pointed out. First, we were intentionally vague about what the output and postselection registers actually consist of, since these depend on the details of model A. If A corresponds to classical (resp.\ quantum) circuits, then  $o$ and $p$ represent subsets of the output bits (resp.\ qubits), and the success condition corresponds to obtaining some particular bit string on $p$. However, if A corresponds to linear optics, then $o$ and $p$ correspond to a subset of the output modes, and the success condition will correspond to observing some configuration of photons in $p$. The changes in the formal definitions necessary for each of these cases, however, are very straightforward. The second subtlety is in the size of the register $p$. In some cases, such as when defining postBQP (see, e.g., Ref.\ \cite{Aaronson2005}), $p$ is considered to be a single qubit without loss of generality, as it is trivial to encode, on a single qubit, whether poly$(n)$ output qubits are in a particular bit string using the operations available in BQP. However, the same is not true for other models, such as IQP, LO or constant-depth quantum circuits, due to the limited set of operations allowed, and there we must consider the most general case where $p$ is a polynomial-sized register. In fact, for these three cases it is possible to show that the probabilities associated with logarithmic-sized subsets of the output register can be computed efficiently classically \cite{Terhal2004, Bremner2011,Aaronson2013a}, and only by looking at sufficiently large outputs can we obtain the models' ``full'' computational power.

Now we invoke a set of well-known results from complexity theory which ultimately provide evidence that, even with the unrealistic power of postselection, classical computers still seem to be much less powerful than quantum computers (for formal definitions of the complexity classes mentioned here, see Refs.\ \cite{Aaronson2013a, Bremner2011}\footnote{Or the Complexity Zoo at \url{https://complexityzoo.uwaterloo.ca/Complexity_Zoo}}). First, it is known that P with a postBPP oracle is contained within the third level of the polynomial hierarchy \cite{Han1997}, which is a recursively-defined tower of complexity classes strongly conjectured to be infinite. Second, a result by Aaronson \cite{Aaronson2005} shows that postBQP is in fact equal to a complexity class known as PP. Finally, a theorem by Toda shows that P with a PP oracle contains the whole polynomial hierarchy \cite{Toda1991}. Concatenating these results one can see that, if postBPP were equal to postBQP, the polynomial hierarchy would collapse to its third level, a surprising result that would contradict widely believed conjectures.

Suppose now that our restricted model A, mentioned above, is such that postA = postBQP. That is, suppose that A with postselection is as strong as BQP (with postselection), even if A is not expected to be equal to BQP itself. Then the previous paragraph proves that an efficient classical weak simulation of A (by which we mean the ability to sample exactly from the output distribution produced by a device from A with only a classical computer and polynomial resources \cite{Bremner2011}) is unlikely to be possible. To see this, note that if we could program a classical computer to efficiently sample from the output distribution of A, then this would mean that postA $\subseteq$ postBPP, since we could just apply the same postselection on the output of the classical simulation that would be applied on the output of A. But, as argued before, this would lead to postBPP = postBQP and the collapse of the polynomial hierarchy.

This relation, postA = postBQP, is in fact obtained for constant-depth circuits of depth greater than three \cite{Terhal2004}, for IQP \cite{Bremner2011}, for MBCC \cite{Hoban2014} and for LO \cite{Aaronson2013a} (as we will review on the next section), and is the crucial reason why all of these restricted models are expected to be hard to simulate classically. In this work, we will show that the exact same relation holds for constant-depth linear optics, thus adding it to this set of restricted models.

One important aspect of this postselection argument is that it only provides evidence that it is hard to sample from the \emph{exact} output distribution of the corresponding device (or at most from a distribution close within multiplicative error to the exact one, which is a very strong requirement given that most probabilities will be exponentially small \cite{Aaronson2013a}). In this sense, BosonSampling seems to be more robust than the other models: the main contribution of Ref.\ \cite{Aaronson2013a} was to prove that, under some reasonable technical conjectures, this sampling should be classically hard even in an \emph{approximate} sense (more specifically, even if we change the task to sampling from any distribution close to the ideal one in total variation distance). This result, however, relies on different techniques than the postselection argument described and used here.

\subsection{The KLM scheme and BosonSampling} \label{sec:KLM}

In 2001 a seminal work by Knill, Laflamme and Milburn \cite{Knill2001b} showed how to implement any quantum circuit in a scalable manner using only linear-optical elements and adaptive measurements. Several variants have been developed since then, with the purpose of reducing the large (but polynomial) overheads in resources induced by the original construction. Here we will limit our discussion to that variant which is more suitable for our purposes, but for a full review see e.g.\ Ref.\ \cite{Kok2007}.
   
The KLM scheme uses what is known as dual-rail encoding, where each qubit is encoded using one photon and two modes. Each computational state corresponds to the photon being found in one of the modes, as such:
\begin{align}
\ket{0}_L = \ket{10},  \notag \\
\ket{1}_L = \ket{01}. \label{eq:KLMencoding}
\end{align}

\begin{figure}[t]
\centering
\subfloat[]{\centering \includegraphics[width=0.25\textwidth]{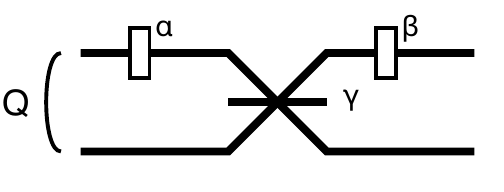}} \\
\subfloat[]{\centering \includegraphics[width=0.4\textwidth]{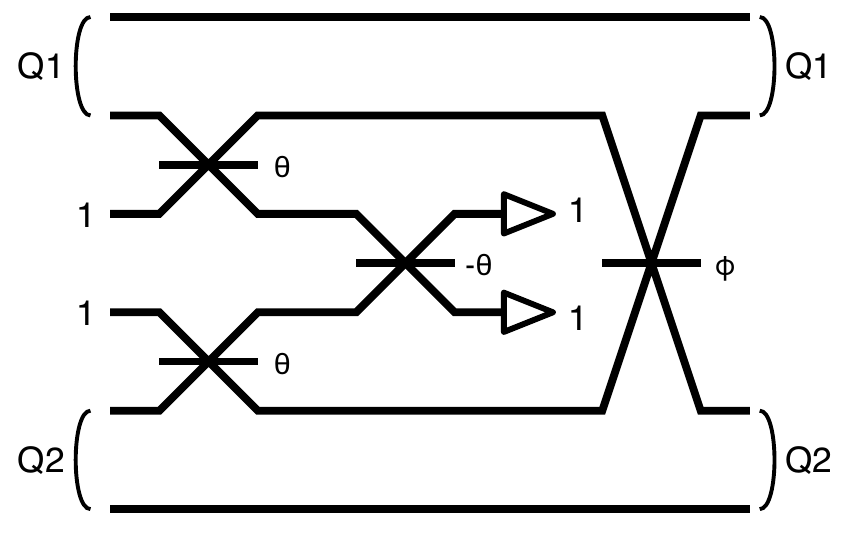}} 
\caption{The KLM scheme. (a) An arbitrary single-qubit gate, up to a global phase, on qubit Q. (b) A probabilistic two-qubit gate. Q1 and Q2 encode the qubits, ordered such that photons on the innermost modes correspond respectively to states $\ket{1}_L$ of each qubit. The two central modes are ancilla modes, initialized each with a single photon. If $\phi \approx 17.63^{\circ}$ and $\theta \approx 54.74^{\circ}$,  and one photon is measured in the output of each ancilla mode, the overall action of the circuit is a $\cz$ gate.}
\label{fig:KLMscheme}
\end{figure}

It is easy to see that, in this encoding, any single-qubit gate can be implemented deterministically using an arbitrary two-mode transformation, as represented in \sfig{KLMscheme}{a}.

To obtain a universal gate set, we also need an entangling logical two-qubit gate \cite{LivroNielsen}. The simplest candidate is the $\cz$ gate, since it only acts nontrivially on the $\ket{11}_L$ state, thereby only requiring two of the four modes involved in the two-qubit state to interact. It suffices to implement some two-mode transformation that adds a $-1$ phase only if both modes are occupied by a single photon, and does nothing otherwise. This seems to require an interaction between the photons: informally, the physical transformation would affect one photon conditioned on the other photon ``being there''. The crucial point of the KLM scheme is that this interaction can be simulated by a measurement, at least probabilistically. In \sfig{KLMscheme}{b} we depict an optical circuit due to Knill \cite{Knill2002} which does precisely that. By adding two extra ancilla modes, each carrying a single photon, and upon measuring again one photon in each of these modes, as depicted in \sfig{KLMscheme}{b}, the circuit is successful and implements a $\cz$ gate. This happens with probability 2/27, and any other measurement outcomes results in a failure of the gate. If this was the only way to implement such transformations then scalable linear-optical quantum computing would be hopeless, as the overall success probability of any circuit would decrease exponentially with the number of two-qubit gates. The original KLM paper also included an error-correcting step to deal with this issue, but we do not need to address it here---our interest is in the capabilities of linear optics with postselection, so we can implement any quantum circuit just by postselecting on the success of every gate and never have to worry about what happens when they fail. 

Replacing error-correction with postselection on the KLM scheme already shows that postLO = postBQP and thus, by the discussion in the previous section, provides evidence against a possible classical simulation of linear optics even in the non-adaptive setting. This was pointed out in Ref.\ \cite{Aaronson2013a}, but the authors go much further. They define a random instance of BosonSampling as 

\begin{itemize}
\item[(i)] preparation of $m$ photonic modes, the first $n$ containing one photon each (with $m$ suitably larger than $n$); 
\item[(ii)] evolution of these modes according to an $m \times m$ random interferometer $U$ sampled from the uniform distribution; and 
\item[(iii)] measurement of the output in the coincidence basis (i.e.\ only considering outputs with a single photon per mode). 
\end{itemize}

By relating the outcome probabilities to the permanents of certain sub-matrices of $U$, together with very plausible conjectures about the hardness of computing the permanent for this particular ensemble of random matrices, the authors are able to provide the more robust result mentioned in the previous section without resorting to the postselection argument, as well as provide a worst-case/average-case equivalence conjecture for random BosonSampling instances. 

In the next section we will consider a variation of BosonSampling where condition (ii) is replaced by

\begin{itemize}
\item[(ii$^\prime$)] evolution of these modes according to an $m \times m$ interferometer composed of four layers of arbitrary two-mode transformations;
\end{itemize}

Remark that, in our model, the interferometers are not random, but rather they are constructed in a prescribed manner. Also, from this point on we will only consider linear optical circuits where the inputs consist of each photon in a different mode [i.e.\ condition (i)]. This is justified by the fact that both the KLM scheme and the original BosonSampling model satisfy this condition, but also because initializing several of the $n$ photons in the same mode tends to make the simulation easier \cite{Aaronson2013a}.

\section{Constant-depth quantum computing} \label{sec:cdepthQC}

Let us begin with an alternative proof of the results found in Ref.\ \cite{Terhal2004}. In that paper, the authors show that quantum circuits with depth three can simulate any arbitrary quantum circuit if postselection is allowed---from the discussion of \sec{postselection} this immediately means that an efficient classical weak simulation of such circuits would imply a collapse of the polynomial hierarchy. They also show that this depth is the smallest for which this holds by giving an explicit simulation for circuits of depth $\leq 2$. For the present purposes, the depth of a quantum circuit is defined as the number of layers of arbitrary two-qubit gates (the authors in Ref.\ \cite{Terhal2004} use a different convention, including the final round of measurements in the depth count, but this only changes the definition of depth by one). We consider that gates acting on qubit pairs that share a common qubit (e.g.,  $\{1,2\}$ and $\{2,3\}$) cannot be done simultaneously even if they commute. Single-qubit gates do not contribute to the depth count since we can consider them as part of the nearest two-qubit gate which means, in particular, that we can choose initialization and measurements in any single-qubit basis without loss of generality.

Our starting point is an universal construction in the measurement-based model of quantum computing (MBQC) \cite{Raussendorf2001, Raussendorf2012}. We will restrict ourselves to a description of how the computation is performed, with no discussion on how its universality is proved---that would lie beyond the scope of this work, and we refer to \cite{Broadbent2009} and references therein.

\begin{figure}[t]
\centering
\subfloat[]{\centering \includegraphics[width=0.45\textwidth]{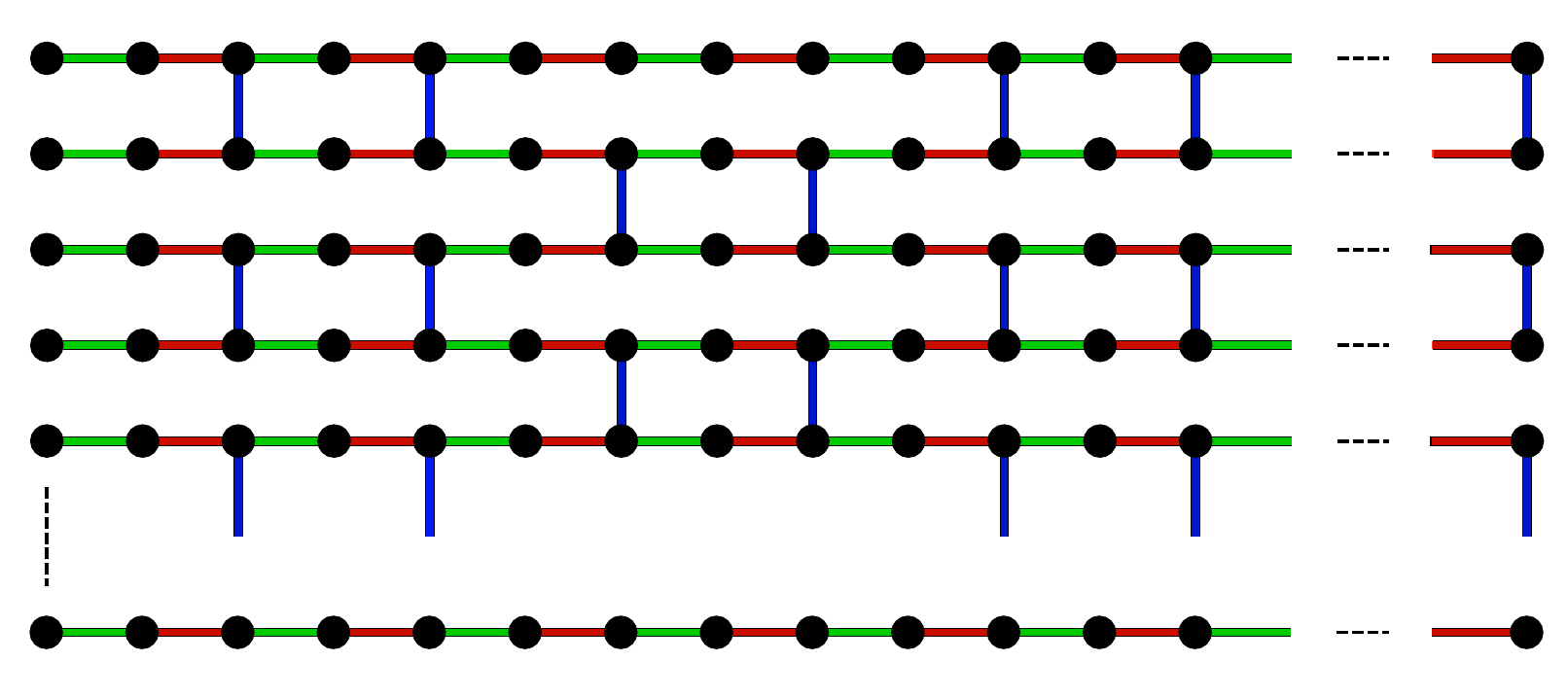}} \qquad
\subfloat[]{\centering \includegraphics[width=0.35\textwidth]{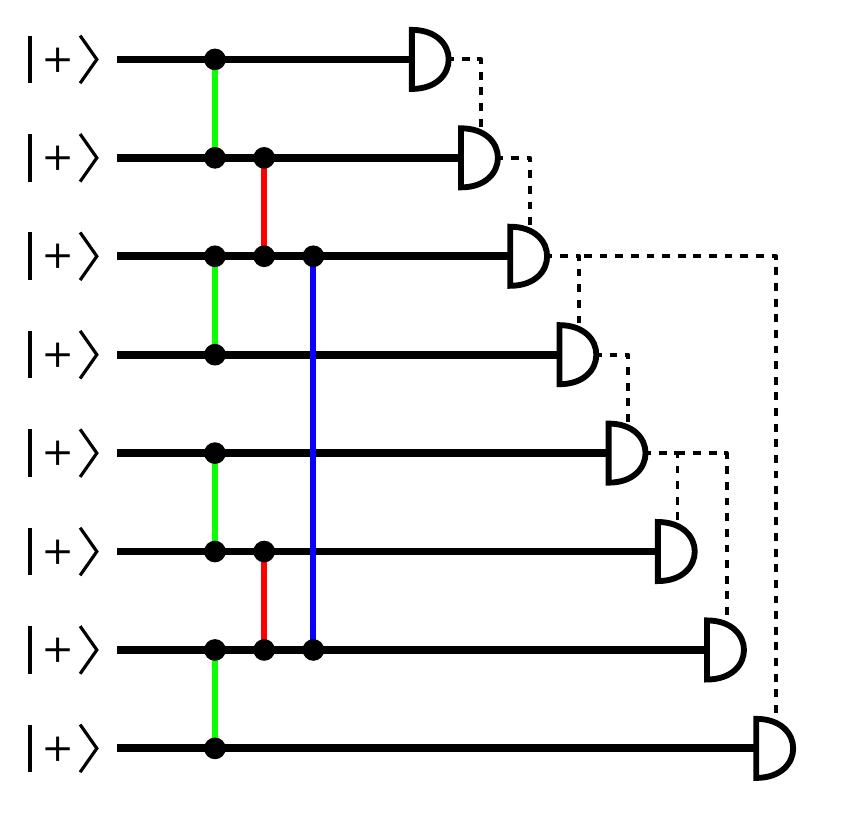}}
\caption[The brickwork graph state.]{(a) The brickwork graph state. (b) Representation of the MBQC protocol as a circuit. Dashed lines conceptually represent the dependence of some measurements bases on the others. The color coding shows how, in the translation from (a) to (b), preparation of the state corresponds to only three rounds of $\cz$ gates, while almost all of the temporal structure resides on the measurement adaptation.}
\label{fig:brickwork}
\end{figure}

Let $G$ be the family of brickwork graphs, such as in \sfig{brickwork}{a}, parameterized by some size parameter $n$ (i.e.\ the graph has poly$(n)$ vertices). Consider, then, the corresponding graph state $\ket{G}$, built as follows: (i) for each vertex in $G$ prepare a qubit in the $\ket{+}$ state and (ii) for each edge in $G$ apply a $\cz$ gate between the two corresponding qubits. This generates a highly-entangled multi-qubit state, and the computation then proceeds by a sequence of single-qubit measurements on the qubits of $\ket{G}$. Each measurement is done in one of a discrete set of bases, and their outcomes determine the bases of future measurements. The complete set of measurements, including the order in which they are performed and the dependence of some measurements on the results of others, is known as a \emph{measurement pattern}. 

We now restate the following Theorem, which is taken from \cite{Broadbent2009} (we follow the result of \cite{Broadbent2009} for convenience, but the brickwork state was known to be universal for MBQC before that, see e.g.\ \cite{Childs2005}):

\begin{theorem} \label{thm:MBQC}
Let $G$ be the brickwork graph shown in \fig{brickwork}. The corresponding graph state is universal for measurement-based quantum computation. The computation is performed as follows:
\begin{itemize}
\item[(i)] Initialize a qubit in the $\ket{+}:=(\ket{0}+\ket{1})/\sqrt{2}$ state for every vertex of $G$;
\item[(ii)] Apply $\cz$ gates according to the edges of $G$;
\item[(iii)] Sequentially measure each qubit of $\ket{G}$ in one the bases 
\begin{equation*}
\left \{\ket{\pm_\theta}=\ket{0} \pm \textrm{e}^{i \theta} \ket{1} | \theta = 0, \pm \pi/4, \pm \pi/2 \right \}.
\end{equation*}
\end{itemize}
Any poly-sized quantum circuit on $n$ qubits can be simulated in this way using some graph $G$ of poly$(n)$ size, and by a suitable choice of the measurement pattern which, furthermore, can be computed efficiently classically.
\end{theorem}

A formal proof of this result, found in Ref.\ \cite{Broadbent2009} and references therein, gives explicit measurement patterns corresponding to gates in the universal set $\{ \cnot, H, T \}$, and thus an explicit procedure to simulate any quantum computation. Omitting further details from this proof, we would just like to point out that this protocol, described as a quantum circuit (cf.\ \sfig{brickwork}{b}), inherits a temporal structure almost exclusively from the ``classical'' adaptation of measurements. In other words, preparation of the graph state only takes a few rounds of two-qubit gates, and the depth of the circuit stems from the fact that some qubits \emph{must} be measured prior to others.

Now consider what happens if we replace adaptive measurements by postselection. That is, rather than making a measurement and conditioning future measurements on its outcome, we just \emph{postselect} each measurement to a given outcome such that adaptation is unnecessary. As an example, suppose we measure one qubit in basis $M_1$, with two possible outcomes labeled $+$ and $-$, and the measurement pattern dictates that a second qubit must be measured either on basis $M_2^{+}$ or basis $M_2^{-}$ depending on the outcome of $M_1$. This can be replaced simply by postselecting $M_1$ to the outcome $+$ and simultaneously measuring the second qubit in basis $M_2^{+}$. Doing this for every measurement effectively flattens out the temporal structure of the protocol, allowing us to perform all measurements in a single round---if this sounds too good to be true, recall that postselection is indeed an extremely unrealistic ``power''. 

Our goal is now basically reached, as the procedure described already constitutes a quantum circuit of depth three. To see this, consider the circuit description of the MBQC protocol, as shown in \fig{brickwork}. In this circuit, the depth count goes as follows: (i) a first round of $\cz$ gates (green edges of \sfig{brickwork}{a}); (ii) a second round of $\cz$ gates (red); (iii) a third round of $\cz$ gates (blue), followed by a round of single-qubit gates to prepare the measurement bases (recall that neither state preparation in the $\ket{+}$ state nor preparation for single-qubit measurements count for the depth). Only three rounds of $\cz$ gates suffice because vertices of the brickwork state have degree at most three, as shown in \fig{brickwork}. It should be clear by our arguments and \thm{MBQC} that circuits of this form, when imbued with the power of postselection, can implement any computation in postBQP. We then conclude, by the previous discussion, that an efficient classical simulation of its output would imply collapse of the polynomial hierarchy.

It is interesting that the MBQC approach is very well suited for this proof, since almost all the temporal structure lies in the adaptive measurements, which is precisely what we replace by postselection. Curiously, all information about the computation \emph{also} lies in the measurements, since the brickwork state does not depend at all on the underlying quantum computation (except, of course, in its size). Thus, it seems that all ``computational power'' (to abuse the terminology) of the constant-depth quantum circuit resides on the combinations of possible choices of measurement bases. An interesting open question is whether there is a natural way to randomly choose the measurement bases such that the final simulation is a particularly hard instance, similar to the worst-case/average-case equivalence conjectured for approximate BosonSampling \cite{Aaronson2013a}.

Another curious aspect of this proof is that it encompasses several similar proofs for different models. The resulting circuit obtained by flattening out the adaptivity of MBQC has depth three, is an MBCC protocol (by definition), and is in IQP, since it only uses gates diagonal in the $X$ basis. This is interesting, but maybe not specially surprising---the concept of gate teleportation is present, in one form or another, in all of these results, and is closely related to the historical origin of MBQC. We should also point out that, while this proof unifies several results, it does not provide concrete relations between these models---the resulting circuit lies at their intersection, but each model may have circuits that perform tasks outside of this intersection.

\subsection{Quantum circuits of depth two} \label{sec:2depthQC}

Besides proving that a depth of three is sufficient for the hardness result, the authors in Ref.\ \cite{Terhal2004} also prove that it is necessary. This is done by giving an explicit simulation of any depth-2 circuit, which we briefly reproduce here.

Consider an arbitrary depth-2 circuit. That is, the qubits are initialized in the computational basis, undergo a first round of arbitrary two-qubit gates, followed by a second round of arbitrary two-qubit gates, and a final round of measurements. The simulation becomes straightforward if we reinterpret this as a two-round computation: first the qubits are prepared in arbitrary two-qubit states, and then they are measured in arbitrary two-qubit bases. The classical simulation then proceeds as follows: 

\begin{itemize}
\item[(i)] Choose one particular measurement $M_i$. This is a two-qubit measurement where each qubit may be entangled with some other qubit. Thus, its outcome probabilities only depend on a four-qubit state and are trivially easy to compute in constant time.
\item[(ii)] Compute the outcome probabilities and simulate $M_i$ by classically sampling a two-bit string from the corresponding distribution and fixing the output of the measurement accordingly.
\item[(iii)] Project the two measured qubits onto the sampled bit string, and update the description of the two qubits entangled to them accordingly. These two qubits are now in an arbitrary, but known, two-qubit state.
\item[(iv)] The next measurement involving one of the qubits updated in (iii) is again a two-qubit measurement that depends only on an arbitrary four-qubit state. 
\item[(v)] Repeat this procedure until all measurements have been fixed.
\end{itemize}

This simulation samples from the same distribution as the corresponding quantum circuit. It is also an example of weak simulation since at no point we compute the probabilities associated with the full output of the circuit, only small two-bit strings at a time. As a consistency check, one could attempt to generalize this simulation for circuits of depth three. This would naturally fail, because the first measurement in step (i) would depend on an 8-qubit state, hence step (iii) would collapse the unmeasured qubits to an arbitrary 6-qubit state. But then, the next time we returned to step (i) the measurement might depend on the state of 14 qubits and so on, and the iteration is clearly disrupted.

\section{Constant-depth linear-optical circuits} \label{sec:cdepthLOQC}

We now combine the previous results into one---the computational complexity of exactly simulating constant-depth BosonSampling. 

The first step is a rather straightforward concatenation of the results mentioned so far. Consider a constant-depth circuit obtained from some universal MBQC protocol by replacing adaptive measurements with postselection, as in \sec{cdepthQC}. Now map the resulting quantum circuit to a linear-optical circuit using the constructions of \fig{KLMscheme}. Finally, replace the adaptive measurements of the KLM protocol by postselection. The resulting circuit is naturally as strong as postBQP, and thus it cannot have an efficient classical simulation unless the polynomial hierarchy collapses. 

Let us now count the depth of the resulting optical circuit. The single-qubit gates cannot be ``absorbed'' into the two-qubit gates anymore, as was done in \sec{cdepthQC}, since in the linear-optical setting we have a dual-rail encoding where both single- and two-qubit gates correspond to two-mode transformations (i.e.\ the single qubit gates involve the two modes of a single qubit, while two-qubit gates involve two modes of different qubits). We will, however, absorb phase shifters into the closest beam splitter, and count layers of arbitrary two-mode transformations. The total count then goes as follows: one layer of balanced beam splitters for the preparation of the $\ket{+}_L := (\ket{0}_L+\ket{1}_L)/\sqrt{2}$ states, followed by six layers for the preparation of the entangled state (i.e.\ two for each layer of $\cz$ gates of the brickwork state, using the $\cz$ of \sfig{KLMscheme}{b}). Finally, one last layer of beam splitters and phase shifters for the preparation of measurement bases, and a round of measurements. This amounts to a total depth of eight.

We can now use two tricks to reduce this depth count. First note that, according to \sfig{KLMscheme}{b}, when we perform a $\cz$ gate, only one mode of each encoded qubit is involved, while the other remains idle. By a simple adaptation of the circuit of \sfig{KLMscheme}{b}, we can use the second mode of each qubit to simultaneously implement the second round of $\cz$s. To do this, we include two $\pi$ phase shifters before the circuit of \sfig{KLMscheme}{b}, resulting in a gate which, up to a global phase, only adds a minus sign to the photonic $\ket{00}$ state. It is clear that, if we act with this gate on the modes encoding the $\ket{0}_L$ state of two qubits, we obtain precisely a logical $\cz$ gate. By a clever alternation of which mode we use for each $\cz$ gate, we can build the long paths of the brickwork state (i.e.\ the red and green edges in \fig{brickwork}) using one round of $\cz$s, corresponding to two rounds of beam splitters. This reduces the overall depth to six.

\begin{figure}[t]
\centering
\subfloat[]{\centering \includegraphics[width=0.45\textwidth]{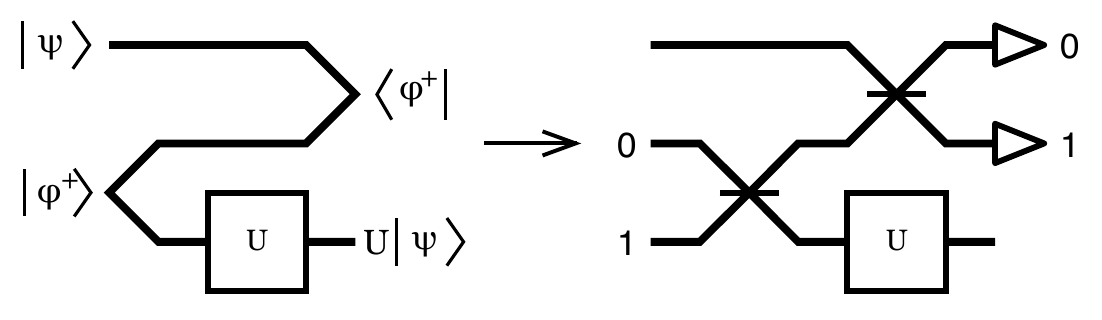}}\\
\subfloat[]{\centering \includegraphics[width=0.45\textwidth]{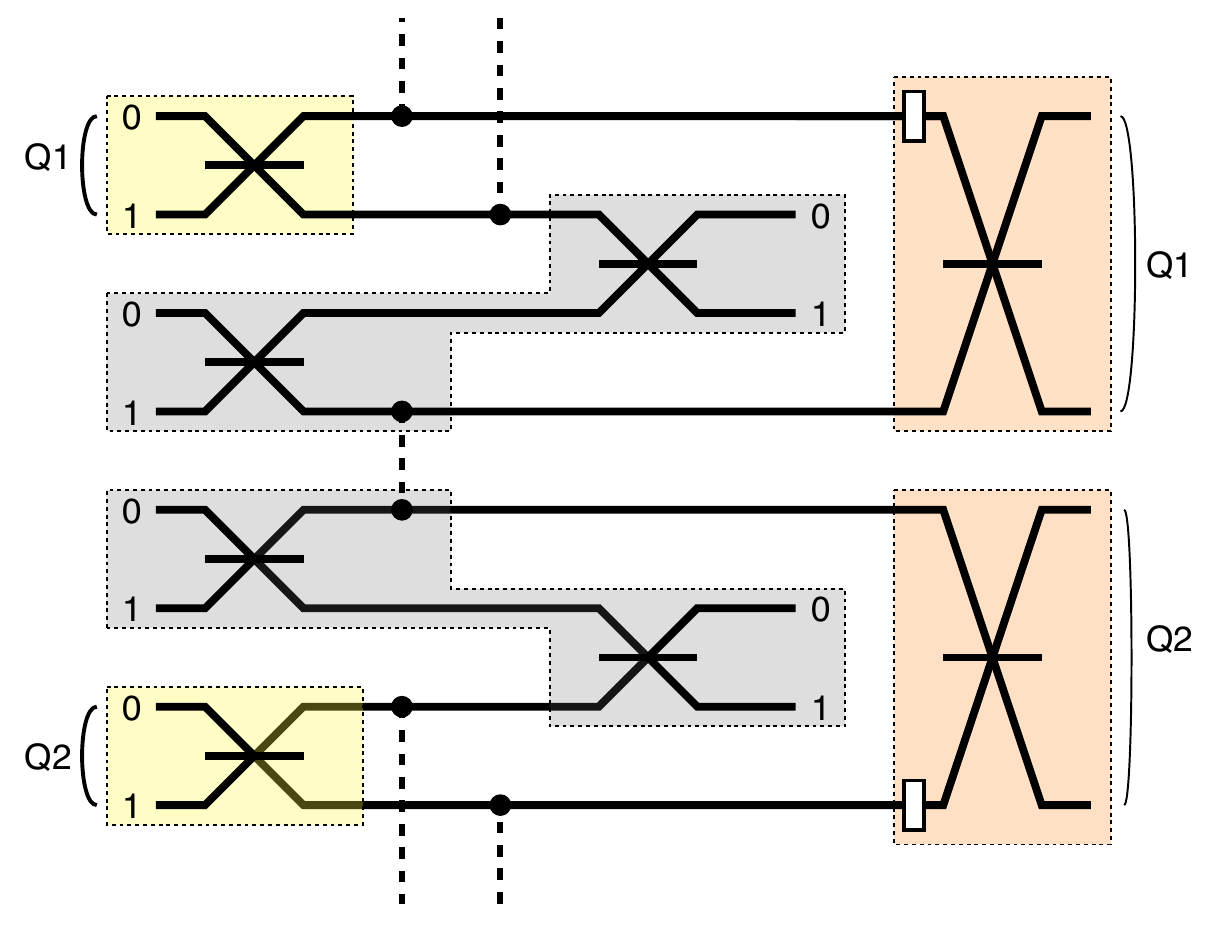}}
\caption{(a) Standard gate teleportation in the quantum circuit model and its linear-optical version, where the two-mode state $\frac{1}{\sqrt{2}}(\ket{01}+\ket{10})$ is mapped to the Fock basis by a balanced beam splitter. (b) The resulting constant-depth linear-optical circuit. Every beam splitter in this figure is balanced, and white rectangles represent phase shifters. Dashed lines represent the probabilistic $\cz$ gates of the usual KLM scheme (i.e.\ the four central modes in \sfig{KLMscheme}{b}), shown implicitly so as to not clutter the figure. The color coding is intended to aid in the mapping from the construction of \sec{cdepthQC}: yellow regions represent state preparation, gray regions represent gate teleportation, and orange regions represent single-qubit measurements. The depth bottleneck of the model is seen here in the two central modes, each involved in four beam splitters (two explicit and two implicit).}
\label{fig:teleportation}
\end{figure}

The second trick is to use a procedure similar to the usual gate teleportation \cite{Gottesman1999b}, which can be found in Ref.\ \cite{Knill2001b}, to parallelize the third layer of $\cz$s. Say we want to implement some faulty (i.e.\ probabilistic) gate $U$ on a computational qubit, but without endangering the information stored in that qubit. We can then do the following: we prepare an auxiliary two-qubit state $\ket{\phi^+}=\frac{1}{\sqrt{2}}(\ket{01}+\ket{10})$, implement $U$ on the second qubit of this auxiliary state, and, if $U$ succeeds, project the first qubit of the auxiliary state together with our original qubit on the $\bra{\phi^+}$ state (see \sfig{teleportation}{a}). Of course, in standard quantum computation (i.e.\ BQP) we cannot guarantee that a two-qubit measurement will project our qubits on the $\bra{\phi^+}$ state, but for our purposes we can just postselect on observing this outcome. Now notice that, if rather than being a two-qubit state, $\ket{\phi^+}$ actually represents a state of one photon in two modes, the mathematical structure is exactly the same, and so it allows us to use an equivalent scheme to teleport the states of the \emph{modes} in the KLM scheme. Since each $\cz$ gate only involves one mode of each qubit, as in \sfig{KLMscheme}{b}, this allows us to implement the gate teleportation by teleporting only one mode, rather than a complete qubit, as shown in \sfig{teleportation}{a}. Thus, we can perform all the $\cz$ gates for the brickwork state in parallel, and just teleport the computational states around, in a very similar spirit to the original result of \cite{Terhal2004}. This scheme, and how it can used to reduce the depth, is illustrated in \sfig{teleportation}{b}. This reduces the total depth count to four.

Note that, of the two tricks mentioned, the second would already suffice to reduce the depth count to four. We presented the first as it may be of independent interest for the parallelization of linear-optical circuits even in future implementations of the full KLM scheme. Also, since in this setting we do not have to worry about the gates failing, it would be wasteful to do all the $\cz$'s in fresh ancilla modes rather than directly on the qubit themselves whenever possible.

\subsection{Depth-2 linear optics} \label{sec:2depthLOQC}

We can now also ask whether, besides being sufficient, a depth of four is also necessary for the preceding result. Alas, unfortunately so far the answer seems to be no. A direct adaptation of the simulation of \sec{2depthQC} only proves that depth-2 linear optical circuits are classically simulable.

Our simulation follows the same basic principle as that described in \sec{2depthQC}, namely that we can re-interpret a depth-2 linear optical circuit as a two-mode preparation followed by a two-mode measurement. The main difference between the simulation of \sec{2depthQC} and the analogous for linear optics is the following: assuming that each mode is initialized with either zero or one photons (i.e.\ $\ket{0}$ and $\ket{1}$), the state after the first layer of beam splitters can have some modes occupied by two photons (i.e.\ $\ket{2}$). Thus one could argue that, when we do the iterative step of the simulation, the sample space changes from one step to the next, possibly disrupting the iteration. But, if we observe that no mode can ever have more than four photons, this can easily be accommodated into the simulation scheme as follows:

\begin{itemize}
\item[(i)] Choose a measurement $M_i$. This is a two-mode measurement, where each mode has up to two photons and may be ``entangled'' with some other mode (by which we mean ``mode entanglement'', where two modes are written in a nonseparable state such as $(\ket{01}+\ket{10})/\sqrt{2}$, not to be confused with ``real entanglement'' between two photons due to some interaction). Thus, this measurement only depends on the states of four modes, each with up to two photons, and its probabilities can be computed in constant time.
\item[(ii)] Compute the outcome probabilities and simulate $M_i$ by sampling from the corresponding distribution and fixing the output of the measurement accordingly. Note that the space of outcomes can have from zero to four photons distributed in any way among the two measured modes.
 \item[(iii)] Project the two measured modes onto the sampled state, and update the description of the two unmeasured modes accordingly. They are now in an arbitrary, but known, two-mode state. By photon-number preservation, the outcome fixed in (ii) determines the number of photons in the collapsed state, but it is easy to check that it can have at most two photons per mode.
\item[(iv)] The next measurement involving one of the modes collapsed in (iii) is again a two-mode measurement that depends only on a four-mode state. One caveat is that these four modes may now contain up to eight photons overall (up to two per mode) depending on previously fixed outcomes. Nonetheless, any subsequent two-mode measurement can only detect four of these photons, so step (ii) can always be applied.
\item[(v)] Iterate this procedure until all measurements have been fixed.
\end{itemize}

As should be clear, since we have only two layers of beam splitters, no mode can ever have more than two photons, except immediately prior to a measurement, in which it can have up to four photons. This guarantees that, even though the first measurement differ from subsequent measurements in the set of allowed states, this does not disrupt the iterative procedure. Also note that this simulation seems to break down for more layers. If the optical circuit has depth three, say, then a first measurement could depend on the state of eight modes, and collapse the unmeasured modes into an arbitrary state of six modes. But then the next measurement involving these modes could depend on 14 modes, and so forth. Since each new choice of measurement may depend on a larger number of modes, the iteration step fails.

\section{Final Remarks} \label{sec:conclusion}

In this paper, we showed that the exact version of the BosonSampling result in Section 4 of Ref.\ \cite{Aaronson2013a} already holds if the linear optical circuit has depth four, and also that it cannot hold for depth two or lower. This seems to be a consequence of the fact that a probabilistic linear-optical $\cz$ gate must be constructed with at least two layers of beam splitters. This is natural, since it is known that ancilla modes and photons are necessary \cite{Knill2002} for the construction of the gate (even probabilistically), and so a single layer of beam splitters would clearly not suffice to interact both the two computational modes and the ancilla modes in question. It cannot be ruled out, of course, that an ingenious use of postselection might provide a construction for a hard-to-simulate depth-3 exact BosonSampling instance via completely different means.

Besides the corresponding result for depth three, we leave another open question: is it possible to somehow close the depth gap between this result and the main result of Ref.\ \cite{Aaronson2013a} for approximate BosonSampling? In the latter case, the minimum necessary depth is O($n$ log $m$), for $n$ photons in $m$ modes, as shown in Ref.\ \cite{Aaronson2013a}, and for exact BosonSampling the minimal known depth, as we showed, is four. There are two natural routes to close this gap: either prove that the approximate BosonSampling result can be simplified, using a further parallelized scheme or possibly a matrix ensemble other than the uniform one, or give an efficient classical algorithm for approximate constant-depth BosonSampling (thus showing that constant-depth and regular BosonSampling are indeed fundamentally different). The latter would not be too surprising: there exist problems for which the exact solution is hard (e.g.\ in $\#$P), while an approximate solution is easy (i.e.\ in P), such as the permanent of positive matrices \cite{Aaronson2013a,Jerrum2004}. Another possibility would be to explicitly investigate the permanents of constant-depth interferometers. However, the unitary of a depth-4 interferometer is at most $16$-sparse, simply because a photon that enters one mode cannot leave in more than $16$ different modes, and similarly a photon that exits in a particular mode cannot have entered in more than $16$ different input modes. Thus, these matrices obey a very special structure, very different from the uniform ensemble, and a worst-case/average-case equivalence allowing us to condition the hardness of their permanents on some plausible conjecture seems less likely.

An approximate result for constant-depth BosonSampling would also have important consequences for the corresponding complexity classes. For example, it can contribute to the more general program of relating the different restricted models described so far. It is known that an arbitrary $n$-photon, $m$-mode linear-optical circuit can be simulated in BQP simply by treating each mode as an $n$-level system, and using a standard mapping from $m$ qudits to O($m$ log $n$) qubits \cite{Aaronson2013a,LivroNielsen}. As discussed in Ref.\ \cite{Terhal2004}, this mapping induces an O(log $n$) depth increase, simply because an arbitrary single-qudit gate will map to an arbitrary $($log $n)$-qubit circuit. This logarithmic depth increase suggests that arbitrary linear-optics might not be simulable by constant-depth quantum circuits. However, the same is not true for constant-depth linear optics. In an optical circuit consisting of a no-collision Fock state input (e.g.\ $\ket{1 1 1 \ldots 1 0 \ldots 0}$) followed by a constant number of beam splitter layers, it is easy to see that no mode can have more than a constant number of photons. For example, if the optical circuit has depth four, no mode ever has more than 16 photons, which means that this system can be viewed simply as a collection of $m$ 16-level systems. Consequentially, simulating it on a quantum computer does not induce a logarithmic depth increase and it can, in fact, be simulated in constant depth. So we see that constant-depth linear optics is provably contained within constant-depth quantum computing (although the second constant might not be three, as in \sec{cdepthQC}), and so showing that a classical simulation of the former is unlikely even in the approximate scenario, along the lines of \cite{Aaronson2013a}, might suggest an equivalent result for the latter. This would be the first evidence that other recent ``collapse-the-polynomial-hierarchy'' results might in fact be as robust as BosonSampling.

Finally, we would like to discuss possible implications of this result for current experimental efforts. To our knowledge, all BosonSampling experiments so far have been performed using integrated photonic circuits \cite{Broome2013, Crespi2013b,Spring2013,Tillmann2013,Spagnolo2013b,Spagnolo2014,Carolan2014}. In these devices, the depth of the circuit is one of the leading causes of photon loss and exponential attenuation of the signal, so any optimization of the model in terms of depth should be beneficial for the experimental efforts. One drawback of the proof given here, however, is that it requires beam splitters acting between arbitrarily distant modes, a requirement that is not as well-suited for usual two-dimensional layouts of these devices. We believe our result may pave the way for practical simplifications of current experimental efforts, but we leave for future work to determine the exact trade-off between losses induced by the depth count and losses induced by having to engineer crossovers between the waveguides.

\begin{acknowledgments}
The author would like to thank Andrew Childs and Ernesto Galv\~ao for helpful and insightful discussions. This research was supported in part by Brazilian funding agency CNPq (Conselho Nacional de Desenvolvimento Cient\'ifico e Tecnol\'ogico), and in part by the Perimeter Institute for Theoretical Physics. Research at Perimeter Institute is supported by the Government of Canada through Industry Canada and by the Province of Ontario through the Ministry of Research and Innovation.
\end{acknowledgments}


\end{document}